\documentclass[aoas,nameyear,dvips]{arximspdf}
\usepackage{multirow}
\usepackage{graphicx}

\doi{10.1214/10-AOAS411}
\volume{5}
\issue{2A}
\pubyear{2011}
\firstpage{873}
\lastpage{893}



\begin{document}
\begin{frontmatter}

\title{Sparse Partitioning: Nonlinear regression with binary or
tertiary predictors, with application to association studies}

\runtitle{Sparse Partitioning}

\begin{aug}
\author[A]{\fnms{Doug} \snm{Speed}\corref{}\ead[label=e1]{doug.speed@ucl.ac.uk}\thanksref{tt1}}
\and
\author[B]{\fnms{Simon} \snm{Tavar\'e}\ead[label=e2]{st321@cam.ac.uk}\thanksref{t2}}

\thankstext{tt1}{Supported by an EPSRC doctoral fellowship.}
\thankstext{t2}{Supported in part by NIH ENDGAME Grant U01 HL084706
and NIH Grant P50 HG02790.}

\runauthor{D. Speed and S. Tavar\'e}
\affiliation{University of Cambridge}
\address[A]{Department of Applied Maths\\
\quad and Theoretical Physics\\
Centre for Mathematical Sciences\\
University of Cambridge\\
Wilberforce Road\\
Cambridge CB3 OWA\\
UK\\
\printead{e1}} 
\address[B]{Department of Oncology\\
Li Ka Shing Centre\\
University of Cambridge\\
Robinson Way\\
Cambridge CB2 ORE\\
UK\\
\printead{e2}}
\end{aug}

\received{\smonth{12} \syear{2009}}
\revised{\smonth{10} \syear{2010}}

%
\begin{abstract}
This paper presents \textit{Sparse Partitioning}, a Bayesian method for
identifying predictors that either individually or in combination with
others affect a response variable. The method is designed for
regression problems involving binary or tertiary predictors and allows
the number of predictors to exceed the size of the sample, two
properties which make it well suited for association studies.

\textit{Sparse Partitioning} differs from other regression methods by
placing no restrictions on how the predictors may influence the
response. To compensate for this generality, \textit{Sparse
Partitioning} implements a~novel way of exploring the model space. It
searches for high posterior probability partitions of the predictor
set, where each partition defines groups of predictors that jointly
influence the response.

The result is a robust method that requires no prior knowledge of the
true predictor--response relationship. Testing on simulated data
suggests \textit{Sparse Partitioning} will typically match the
performance of an existing method on a data set which obeys the
existing method's model assumptions. When these assumptions are
violated, \textit{Sparse Partitioning} will generally offer superior
performance.
\vspace*{3pt}
\end{abstract}

%
\begin{keyword}
\kwd{Sparse Bayesian modeling}
\kwd{nonlinear regression}
\kwd{association studies}
\kwd{large $p$ small $n$ problems}.
\end{keyword}

\end{frontmatter}

\section*{Introduction}

In recent years association studies have surged in popularity, driven
by the ability to interrogate the genome in ever-increasing detail
[\citet{mccarthy}]. The common aim of these studies is to detect
genomic variants that are linked with a particular phenotype. It is
hoped that detecting such variants will bring us closer to
understanding the biological processes at work.

So far these studies have had mixed results. While variants with strong
effects are picked up fairly readily [e.g., \citet{wellcome}], there is
speculation that more subtle associations are being missed
[\citet{cordell}]. This suggests the need to develop more sophisticated
tools that are able to explore beyond the obvious
[\citet{stephbald}].

Formally, an association study can be viewed as a regression problem
consisting of $n$ data points (the samples) and $N$ predictors (the
variants). In this paper we consider the case when each predictor takes
either two or three unique values. This is common in association
studies. For example, a predictor might record presence or absence of a
mutation, or whether a~variant is in a neutral, amplified or deleted
state. We also allow for ``large~$p$, small $n$ problems'' in which the
number of predictors exceeds the sample size. Again, this is often the
case with association studies, owing to the abundance of genetic
variants available to examine.

Currently available regression tools can be characterized by how they
permit predictors to influence the response. For example, many fit an
additive model, which overlooks the possibility that interactions
between predictors might affect the response. The methods which permit
interactions will generally specify the type of interactions they
allow. A key factor affecting performance is whether the data set being
examined conforms to the restrictions the method imposes.
\textit{Sparse Partitioning} tries to avoid placing restrictions on the
underlying model relationship. This should enable it to maintain power
in scenarios where other methods might fail.

Section~\ref{sec:existing} describes some of the existing methods
suitable for processing high-dimensional data.
Sections~\ref{sec:partition} and~\ref{sec:spmethod} briefly outline the
\textit{Sparse Partitioning} methodology. Sections~\ref{sec:simstudy}
and~\ref{sec:real} test the performance of \textit{Sparse Partitioning}
compared to existing methods, while Section~\ref{sec:discussion}
concludes the paper. Additional details are provided in the
\hyperref[app]{Appendix}
and supplementary material.\looseness=-1

\section{Existing methods}\label{sec:existing}

The task of a regression method is to infer how the predictors
influence the response. Let the vector $\mathbf{Y}$ (size $n\times1$)
contain the response values and the matrix $\mathbf{X}$ (size $n \times
N$) contain the predictors. For the $i$th data point, $Y_i$ denotes its
response, while $X_{i,1},\ldots, X_{i,g},\ldots,X_{i,N}$ denote its
predictor values. If we write the regression model as
$l(\mathbb{E}(\mathbf{Y})) = f(\mathbf{X})$, where $l$ is a specified
link function, the aim is to deduce properties of $f(\mathbf{X})$, the
``underlying relationship.'' In particular, we wish to identify the
subset of predictors that contribute toward $f(\mathbf{X})$.


Consider writing the underlying relationship as
\[
f(\mathbf{X}) = f_1(X_{G_{1,1}},\ldots,X_{G_{1,s_1}}) + \cdots+
f_K(X_{G_{K,1}},\ldots,X_{G_{K,s_K}}).
\]
Under this representation, $f(\mathbf{X})$ is influenced by additive
contributions from groups of interacting predictors. $f_k$ describes
the contribution of predictors $G_{k,1},\ldots, G_{k,j},
\ldots,G_{k,s_k}$ to $f(\mathbf{X})$. In this paper, additivity is not
considered\vadjust{\eject} an interaction. Therefore, the predictors in each group are
said to interact with each other, but not to interact with a predictor
in a different group. For the most general relationship, all predictors
feature in one group. In practice, however, we suspect $f(\mathbf{X})$
is far simpler.

We have distinguished existing methods based on two features of their
underlying relationships: whether they permit more than one group of
predictors to contribute to $f(\mathbf{X})$ and whether they permit
interactions between contributing predictors.
Figure~\ref{fig:existingtable} demonstrates the four possibilities,
using the case when the predictors are binary and the response is
continuous.

\subsection{One group, maximum group size one}\vspace*{-2pt}
\[
f(\mathbf{X}) = f_1(X_{G_{1,1}}).\vspace*{-2pt}
\]
The simplest assumption supposes the response is influenced by only one
predictor. Most methods in this category are equivalent to performing
a~maximum likelihood test comparing a null hypothesis, $f(\mathbf{X}) =$
constant, with an alternative, $f(\mathbf{X}) = f_1(X_g)$. \textit{Single} is our implementation of such a method. Considering that these
methods can only detect an associated predictor by its marginal effect,
they are surprisingly successful. They are also extremely fast to run
and therefore very popular [e.g., \citet{stranger}].

\begin{figure}
\begin{center}
\begin{scriptsize}
\begin{tabular}{@{}ccccc@{}}
&\multirow{4}{0.5pt}[4.8pt]{\rule{0.5pt}{92pt}}& \textbf{ONE GROUP} &\multirow{4}{0.5pt}[4.8pt]{\rule{0.5pt}{92pt}}& \textbf{MULTIPLE GROUPS} \\
&&\textbf{OF PREDICTORS} && \textbf{OF PREDICTORS}\\
\hline
\begin{tabular}{@{}c@{}}
\textbf{NO}\\
[1ex]
\textbf{INTERACTIONS}\\
\end{tabular}
&&
\begin{tabular}{@{}c@{}}
$\mathbf{Y} = \alpha+ \beta X_g$\\
[1.5ex]
e.g.,\ \textit{Single}\\
\end{tabular}
&&
\begin{tabular}{@{}c@{}}
$\mathbf{Y} = \alpha+ \sum_1^N\beta_gX_g$\\
e.g., \textit{SSS}\\
\end{tabular}
\\
\hline
\begin{tabular}{@{}c@{}}
\textbf{INTERACTIONS}\\
\end{tabular}
&&
\begin{tabular}{@{}c@{}}
$\mathbf{Y} = f(X_{G_1}, \ldots, X_{G_s})$\\
[2.5ex]
e.g.,\ \textit{Pairs}, \textit{CART}, \textit{RF}\\
\end{tabular}
&&
\begin{tabular}{@{}c@{}}
$\mathbf{Y} =
f_1(X_{G_{1,1}},\ldots,X_{G_{1,s_1}}) + \cdots+$\\
$f_K(X_{G_{K,1}},\ldots,X_{G_{K,s_K}})$\\
[1ex]
e.g.,\ \textit{Logic}, \textit{MARS}, \textit{Sparse Partitioning}\\
\end{tabular}
\\
\end{tabular}
\vspace*{2pt}
\caption{Regression methods can be categorized according to two
features of their underlying relationship. This table shows the four
possibilities, for the case of binary predictors and a continuous
response. Explanations of the existing methods, \textit{Single},
\textit{Pairs}, \textit{CART}, \textit{RF}, \textit{SSS},
\textit{Logic} and \textit{MARS}, are provided in the main text.}
\label{fig:existingtable}
\end{scriptsize}
\end{center}
\vspace*{-5pt}
\end{figure}

Bayesian alternatives are possible [e.g., \citet{balding}] and useful
if certain predictors are thought \textit{a priori} more likely to be
associated. Otherwise they will generally produce the same results as
classical methods.

\subsection{One group, maximum group size greater than one}\vspace*{-2pt}
\[
f(\mathbf{X}) = f_1(X_{G_{1,1}},\ldots,X_{G_{1,{s_1}}}).\vspace*{-2pt}
\]
Even for very high-dimensional problems ($>$500,000 predictors) it is
possible to test exhaustively all pairwise models [cf.
\citet{marchini}]. The method \textit{Pairs} is our extension of
\textit{Single},
performing a maximum likelihood test for each pair of predictors. While
the method could be extended further to consider three or four way
interactions, this is often infeasible due to computation time.

A second method in this category is \textit{CART} [Classification and Regression
Trees; \citet{carttrees}]. \textit{CART} differs from \textit{Pairs} in not insisting
on the full interaction model for associated predictors. For example, a
\textit{CART} model containing two associated predictors might have
only 3
degrees of freedom, even though there are 4 unique vector values
present. Random Forest [\citet{rf}] offers a stochastic interpretation
of this method, constructing a large number of trees in a quasi-random
fashion and summarizing their properties.

\subsection{More than one group, maximum group size one}\vspace*{-2pt}
\[
f(\mathbf{X}) = f_1(X_{G_{1,1}}) + \cdots+ f_K(X_{G_{K,1}}).\vspace*{-2pt}
\]
This underlying relationship allows more than one predictor to be
causal, but restricts the causal predictors to contributing additively.
When there are more predictors than samples, the standard multiple
regression model will become over-saturated and fail.

The classical solution, adopted by Variable Subset Selection, Lasso and
Ridge Regression [described in \citet{esl}], is to introduce a penalty
term that limits the number of contributing predictors. However, this
penalty term can appear quite arbitrary. An alternative is offered by
Bayesian methods [\citet{wang}; \citet{zhang}; \citet{hoggart}]. These methods allow our
preference for sparse models to be reflected in the prior distribution.
We picked Shotgun Stochastic Search [\citet{sss}] to represent this
category of methods in the simulation studies.

\subsection{More than one group, maximum group size greater than one}\vspace*{-2pt}
\[
f(\mathbf{X}) = f_1(X_{G_{1,1}},\ldots, X_{G_{1,s_1}}) + \cdots+
f_K(X_{G_{K,1}}, \ldots,X_{G_{K,s_K}}).\vspace*{-2pt}
\]
Allowing both interactions and multiple groups of predictors to
contribute to the underlying relationship has the potential of most
accurately describing the true model. However, both decisions increase
the size of the model space and so the difficulty of identifying the
true model.

Logic Regression [\citet{logic}] and Multivariate Adaptive Regression
Splines are two of the few methods in this class. Both methods place
restrictions on the permitted functions which reduce the size of the
model space; \textit{Logic} insists on Boolean operators, while \textit{MARS} uses
products of hinge functions. \textit{Sparse Partitioning} falls into
this category, but places no restriction on the types of functions
allowed.\vspace*{-2pt}

\section{Formulating the regression as a partitioning}
\label{sec:partition}

In order to describe \textit{Sparse Partitioning}'s methodology, it is
convenient to formulate
the regression problem as a search for high scoring partitions.
Consider how the underlying relationship groups predictors:
\begin{eqnarray*}
f(\mathbf{X})&=&f_1(X_{G_{1,1}},\ldots,X_{G_{1,s_1}}) + \cdots+
f_K(X_{G_{K,1}},\ldots,X_{G_{K,s_K}})\\[-2pt]
&=&f_1({X}_{\mathbf{G}_1}) + \cdots+ f_K({X}_{\mathbf{G}_K}).\vspace*{-2pt}
\end{eqnarray*}
The disjoint sets $\mathbf{G}_1, \mathbf{G}_2, \ldots, \mathbf{G}_K$ index groups of
associated predictors. If we let $\mathbf{G}_0$ index the ``null group''---the group of predictors in no way associated with the response---then
$\mathbb{G}= \{\mathbf{G}_0,\mathbf{G}_1,\mathbf{G}_2, \ldots,\mathbf{G}_K\}$ defines a
partitioning of $\{1,2,\ldots,N\}$.

In the representation above, predictors are not allowed to feature in
more than one nonnull group. To avoid this restriction, while at the
same time maintaining a partitioning, the predictor set is expanded to
contain $C$ copies of each predictor and $N$ is increased accordingly.
For the remainder of this paper, we describe the method supposing
$C=1$, then explain the changes required when this is not the case.

A partition can also be described by the vector $\mathbf{I} = (I_1, I_2,
\ldots,I_N)$, where~$I_g$ indicates to which group predictor $g$
belongs. This notation will be useful later on and also reminds us that
the ordering within groups is not important. Figure~\ref{fig:example}
gives an example of a simple partitioning and the underlying
relationship to which it refers.

\begin{figure}
{\fontsize{9.5pt}{11.5pt}\selectfont
\begin{center}
\begin{tabular}{@{}lcr@{}}
\begin{tabular}{@{}rcl@{}}
\rule{0pt}{8ex} $\mathbf{I} $&$=$&$ (\overbrace{1 1}^{\mathbf{G}_1}\overbrace{00}^{\mathbf{G}_0}\overbrace{2}^{\mathbf{G}_2})$\\
[1ex]
$f(\mathbf{X}) $&$=$&$ f_1(X_1,X_2) + f_2(X_5)$\\
\end{tabular}
&
\begin{tabular}{@{}c@{}}
\rule{0pt}{5ex} $\Rightarrow$
\end{tabular}
&
\begin{tabular}{@{}c@{}}
\\
$\!\!\!\!\!$For binary predictors and a continuous response:\\
[1ex] $\mathbf{Y} = \alpha_0 + \alpha_{1,1} X_1(1-X_2) + \alpha_{1,2}
(1-X_1)X_2$\\
$+\,\alpha_{1,3} X_1 X_2 + \alpha_{2,1} X_5$\hphantom{$\alpha_{1,2}(1\,X_1)X_2$}\\
\end{tabular}
\end{tabular}
\end{center}}
%
\caption{An example of a partitioning for a problem
containing five binary predictors (each valued 0 or 1) and a continuous
response.} \label{fig:example} 
\end{figure}

The focus of \textit{Sparse Partitioning} is to determine properties of
the partitioning defined by the underlying relationship. Our main
desire is to identify which predictors are not in the null group.
However, it is also useful to know whether predictors feature in the
same nonnull group, indicating interactions. The advantage of
formulating the problem in terms of partitions is that the model space
is likely too vast to hope to detect accurately the explicit underlying
relationship (i.e., determine $\mathbf{f} = \{f_1, f_2, \ldots, f_K\}$ as
well as~$\mathbb{G}$). Fortunately, we are usually more interested in
detecting which predictors are involved, rather than exactly how they
contribute (the latter can be saved for follow-up analysis).

\section{Sparse Partitioning methodology}
\label{sec:spmethod}

\textit{Sparse Partitioning} is a Baye\-sian methodology, so it follows
the usual steps of deciding a prior, calculating the likelihood, then
computing the posterior distribution of models through use of Bayes
formula:
\[
\mathbb{P}(\mathrm{Model}|\mathrm{Data}) \propto\mathbb{P}(\mathrm{Model})
\times\mathbb{P}(\mathrm{Data}|\mathrm{Model}).
\]
Each model is defined by $\{\mathbb{G},\mathbf{f}\}$, a partition and a
corresponding set of functions. However, $\mathbf{f}$ is considered a
nuisance parameter, so we are interested in determining the marginal
posterior $\mathbb{P}(\mathbb{G}|\mathrm{Data})$.

\subsection{Prior}\vspace*{-2pt}
\[
\mathbb{P}(\mathrm{Model})=\mathbb{P}(\mathbb{G}, \mathbf{f})=\mathbb
{P}(\mathbb{G}
)\times\mathbb{P}(\mathbf{f}|\mathbb{G}).\vspace*{-2pt}
\]
The prior for the partition is based on our belief in $p_g$, the
probability that predictor $g$ is associated with the response.
Therefore, $\mathbb{P}(\mathbb{G}) = \mathbb{P}(\mathbf{I})$ is
constructed such that
$\sum_{\mathbf{I}:I_g \neq0}\mathbb{P}(\mathbf{I})=p_g$. Two partitions containing
the same associations are given equal weight. When multiple copies of
each predictor are permitted, $p_g$ equals the probability that at
least one copy of predictor~$g$ is associated. \textit{Sparse
Partitioning} keeps fixed the values of $p_g$, however, it is
straightforward to allow them to vary if more detailed prior
information is available.

Given $\mathbf{G}_k$, the relevant information of function $f_k$ can be
described by the values it assigns each node (unique vector value) of
$X_{\mathbf{G}_k}$. For the example in Figure~\ref{fig:example}, $\mathbf{f}$ can
be described by $\bolds{\alpha}=
(\alpha_0,\alpha_{1,1},\alpha_{1,2},\alpha_{1,3},\alpha_{2,1})$.
Therefore, the prior on $\mathbf{f}$ is equivalent to a prior on $\bolds{\alpha}$, for which \textit{Sparse Partitioning} uses a multivariate
normal distribution.

\subsection{Likelihood}

The likelihood is determined by the regression model. When the response
is continuous (e.g., a quantitative trait), \textit{Sparse
Partitioning} supposes the residuals are normally distributed. When the
response is binary (e.g., a case-control experiment with
affected and unaffected patients), \textit{Sparse Partitioning} uses a
logit link function. The marginal likelihood is obtained by integrating
across the function parameters:
\[
\mathbb{P}(\mathrm{Data}|\mathbb{G}) = \int_{\mathbf{f}} \mathbb{P}(\mathrm{Data}|\mathbf{f},
\mathbb{G}
)\mathbb{P}(\mathbf{f}| \mathbb{G})\,d\mathbf{f} = \int_{\bolds{\alpha}}
\mathbb{P}
(\mathrm{Data}|\bolds\alpha, \mathbb{G})\mathbb{P}(\bolds\alpha|\mathbb{G})\,d\bolds{\alpha}.\vspace*{-2pt}
\]

\subsection{Posterior}

Explicit calculation of $\mathbb{P}(\mathbb{G}|\mathrm{Data})$ would require an
exhaustive search of the space of partitions, which is infeasible even
for reasonably sized problems. Therefore, \textit{Sparse Partitioning}
uses Markov Chain Monte Carlo (MCMC) techniques to estimate statistics
of this posterior distribution. Within each MCMC iteration, two
sampling stages are used: the first proposes, in turn, a change to each
component of $\mathbf{I}$; the second proposes a change to one element of
$\mathbb{G}$. The bulk of \textit{Sparse Partitioning}'s processing time
is spent sampling from the posterior distribution. Therefore, it is
convenient that the two stages can be parallelized
with an almost linear speed-up.\looseness=-1

Full details of the methodology are provided in
\hyperref[app]{Appendix} and Sections 1, 2 and~3 of the Supplementary
Material [\citet{speedsupp}].

\section{Simulation studies}
\label{sec:simstudy}

In total, ten simulation studies were carried out, designed to test
various aspects of \textit{Sparse Partitioning}'s performance and make
comparisons with existing
methods. This section presents results from the first study. Further
details of the methods used to simulate data and the results from the
remaining nine studies are provided in Section 4 of the Supplementary
Material [\citet{speedsupp}].

Typically, each simulated data set consisted of 100 samples, each of
1000 predictors, three of which were causal for the response. Each
regression method was asked to identify its top three associations and
was then scored by how many causal predictors it correctly identified.
Empirical estimates were obtained by averaging over 100 data sets.

The first study examined the case of (uncorrelated) binary predictors
and a continuous response. Each scenario concentrated on a particular
underlying relationship (Models I, II or III) for a particular
frequency of the causal predictors (0.05, 0.1, 0.2, 0.4 or random). The
first model was designed so that each causal predictor contributed
additively, the second featured a~multiplicative interaction, while the
third involved a~``weird'' interaction (see Table \ref{t1}).

\begin{table}
\caption{The first simulation study considered three different underlying relationships}\label{t1}
\begin{tabular*}{278pt}{@{\extracolsep{\fill}}ll@{}}
\hline
\textbf{Model} & \multicolumn{1}{c@{}}{\textbf{Underlying relationship}}\\
\hline
\phantom{II}I & $ Y=X_1 + 1.5X_2 - 2 X_3$\\
\phantom{I}II & $Y=1.5X_1 \times X_2 + X_3$\\
III&$Y=f(X_1,X_2) + X_3;$\\
&$f(0,0)=0$, $f(1,0)=1$, $f(0,1)=2$, $f(1,1)=-1$\\
\hline
\end{tabular*}
\end{table}

\begin{figure}[b]

\includegraphics{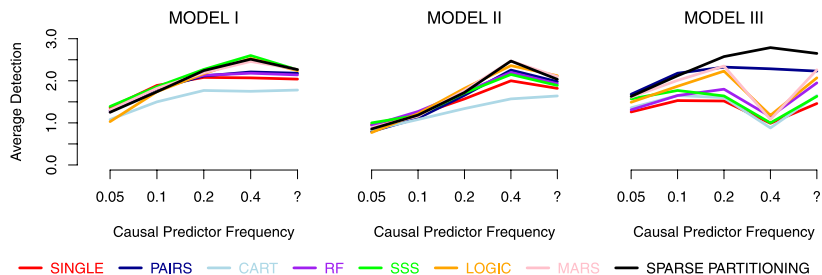}

\caption{Each plot considers a different underlying
relationship (described in the main text). Within each plot, the lines
report the average number of causal predictors correctly detected by
each method for different causal predictor frequencies (``?'' denotes
random).} \label{fig:sim1}
\end{figure}

\begin{figure}

\includegraphics{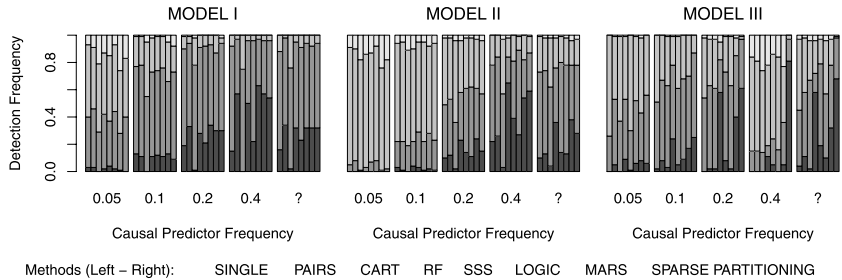}
\vspace*{-3pt}
\caption{Each group of eight vertical bars compares the
methods for a particular underlying relationship and causal predictor
frequency (``?'' denotes random). Within each bar, the lengths of the
shaded sections (from top to bottom) indicate the proportion of time
the method correctly detected 0, 1, 2 and 3 causal predictors. For
example, the lengths of the darkest gray bars show how often each
method successfully identified all three causal predictors. For all
scenarios, the ordering of methods is the same (from left to right):
\textup{Single, Pairs, CART, RF,
SSS, Logic, MARS} and \textup{Sparse
Partitioning}.} \label{fig:sim1b}
\vspace*{-6pt}
\end{figure}

Figure~\ref{fig:sim1} presents results from the first study. Each plot
relates to a different underlying relationship. Within each plot, the
lines display the average number of causal predictors correctly
identified by each method for different frequencies of the causal
predictors. Figure~\ref{fig:sim1b} provides an alternative
interpretation of these results, reporting how often each method
successfully detected 0, 1, 2 or 3 causal predictors for each scenario.

Under Model I, \textit{SSS}, \textit{Logic}, \textit{MARS} and
\textit{Sparse Partitioning} were the four best performing methods
across different frequencies, pulling clear of \textit{Single},
\textit{Pairs} and \textit{RF} as the causal predictor frequency
increased. Under Model II, this order was essentially maintained, with
the exception of \textit{SSS}, which dropped into the second tier of
performers. However, under Model III, \textit{Sparse Partitioning} has
emerged on top, comprehensively beating six of its rivals, with only
\textit{Pairs} coming close.

\textit{Sparse Partitioning} has performed best in this simulation
study, proving itself most robust across the different models. It has
triumphed under~Mo\-del~III, when the underlying relationship assumptions
of all other methods have been violated. Note that its generality has
not prevented it from~at least matching the performance of the existing
methods under Models I and II.\looseness=-1

\section{Application to real data sets}
\label{sec:real}

$\!\!\!$We applied \textit{Sparse Partitioning} to four previously analyzed
association studies: the first looked at expression data~for 109
individuals in the HapMap project
(\href{http://hapmap.ncbi.nlm.nih.gov}{http://hapmap.ncbi.nlm.nih.gov});
the second and third examined
data sets from the ``2010 Project''
(\href{http://walnut.usc.edu/2010}{http://}
\href{http://walnut.usc.edu/2010}{walnut.usc.edu/2010}),
a large-scale study of the plant \textit{Arabidopsis thaliana}; the
fourth used data provided by the Flint laboratory at the University of
Oxford (\href{http://www.well.ox.ac.uk/flint-2}{http://www.well.ox.ac.uk/flint-2}).
Extended versions of
all results are provided in Figures 12--16 of the Supplementary Material
[\citet{speedsupp}].

\subsection{HapMap data}
\label{subsec:hapmap}

\begin{figure}

\includegraphics{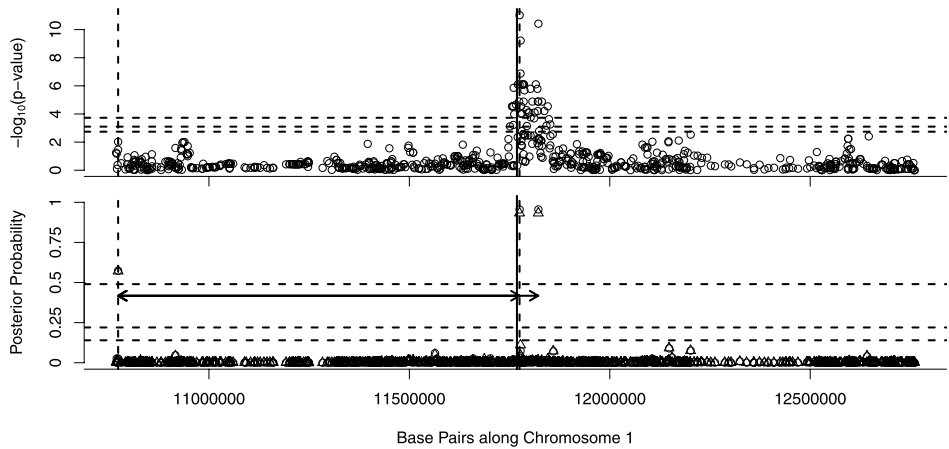}
\vspace*{-3pt}
\caption{Analysis of expression levels of \texttt{MTHFR}
using HapMap data. The top plot shows results from \textup{Single},
the bottom
plot shows results from two runs of \textup{Sparse Partitioning}. Full
details are provided in
the main text.} \label{fig:hapmap}\vspace*{-5pt}
\end{figure}

Dr. Antigone Dimas kindly provided us with a sample of 109 individuals,
each typed for 1,186,075 Single Nucleotide Polymorphisms (SNPs) and
measured for expression levels of 2682 genes [\citet{dimas}]. We
applied \textit{Sparse Partitioning} to the four genes for which Dr.~%
Di\-mas found strongest evidence for an interaction, copying her decision
to consider only SNPs within one million base pairs (1 Mbp) of each
gene. Figure~\ref{fig:hapmap} presents the results for \texttt{MTHFR},
the third of these genes, located approximately 11.8 Mbp along
Chromosome 1. For each SNP in the 2 Mbp region, the top plot displays
the $p$-value obtained by \textit{Single}, while the bottom plot
reports the
posterior probability of association from \textit{Sparse Partitioning}
(circles correspond to run one result, triangles to run two). The solid
vertical line marks the location of the gene, while the two dashed
vertical lines mark the locations of the SNPs declared interacting by
Dr. Dimas. The dashed horizontal lines provide estimates of the 5, 25
and 50\% significance thresholds for the top association of each
method, calculated using permutation tests.\looseness=-1

\textit{Sparse Partitioning} found three promising SNPs, rs2286139,
rs2643888 and rs2279703, with posterior probabilities of association
0.57, 0.96 and 0.96, respectively. It is no coincidence that the second
and third hits have matching probabilities. Before analysis,
\textit{Sparse Partitioning} searches for highly correlated predictors,
as is often the case with fine-scale genetic data. SNP rs2643888 was
found to be highly correlated with SNP rs2279703, with matching values
for 106 of the 109 individuals. Therefore, the former SNP was removed
from analysis, and subsequently given the same posterior estimates as
the latter. \textit{Sparse Partitioning} returned a posterior
probability of interaction of 0.42 between SNPs rs2286139 and
rs2643888/rs2279703 (indicated by the horizontal arrows), offering some
support for Dr. Dimas' findings of an interaction.

\subsection{2010 project: Pilot data}
\label{sub:pilot}

$\!\!$The project's pilot data set looked at 95 accessions, genotyped for
5419 SNPs and measured for ten phenotypic traits. We focused on the
tenth phenotype, expression levels of the \texttt{FRIGIDA} gene. We
decided to remove eight accessions whose genotypes were either almost
identical to remaining accessions or were flagged as suspicious by
principal component analysis. Using methods similar to the original
analysis [\citet{zhao}], we first adjusted the phenotype to correct for
confounding due to population structure and relatedness of accessions.
By contrast, we chose not to impute missing values, meaning
approximately 10\% of the genotypes were supplied to \textit{Sparse
Partitioning} as unobserved.

\begin{figure}

\includegraphics{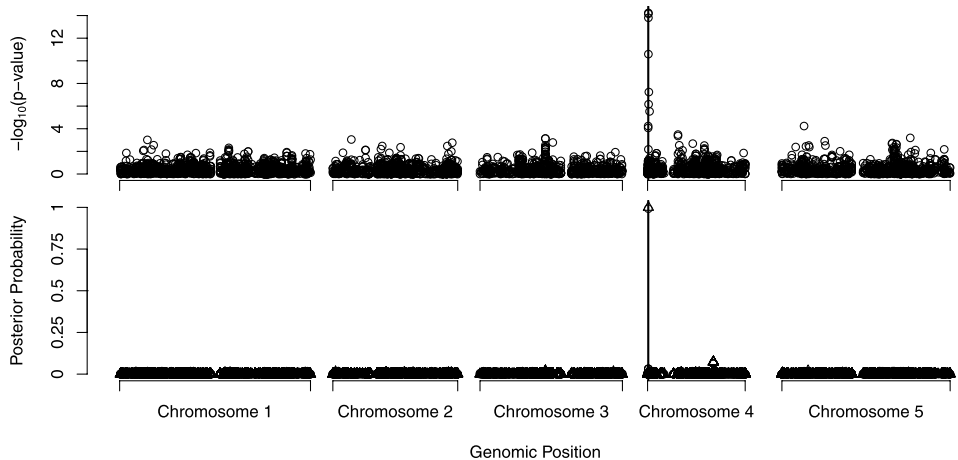}%
\vspace*{-4pt}
\caption{Analysis of expression levels of
\texttt{FRIGIDA} for \textit{Arabidopsis thaliana}. The top plot shows
results from \textup{Single}, the bottom plot shows results from two
runs of
\textup{Sparse Partitioning}. Full details are provided in the main
text.} \label{fig:fri}\vspace*{-6pt}
\end{figure}

Figure~\ref{fig:fri} compares the $p$-values obtained from \textit{Single} to
the posterior probabilities of association of \textit{Sparse
Partitioning}. Our method identified just one strong association,
coinciding with the third strongest hit of \textit{Single} and
suggesting that,
in this case, the simple underlying relationship of \textit
{Single} might be
appropriate. For both methods the strong associations lay very close to
the \texttt{FRIGIDA} region, marked by a solid vertical line,
suggesting the results are accurate.

A possible concern is that \textit{Sparse Partitioning}'s generality
might lead to overfitting on occasions when simple models are more
appropriate. Here that does not appear to be the case, with
\textit{Sparse Partitioning} declaring only one strong association. We
repeated the analysis using imputed data, which allowed us to compare
the prediction accuracy of each method via leave-one-out
cross-validation. The linear model containing only the top hit from
\textit{Single} explained 44\% of the variance, agreeing closely with
\textit
{Sparse Partitioning}'s estimate of 42\% variance explained.\vspace*{-3pt}

\subsection{2010 project: Release 3.04}
\label{sub:release3}

\begin{figure}

\includegraphics{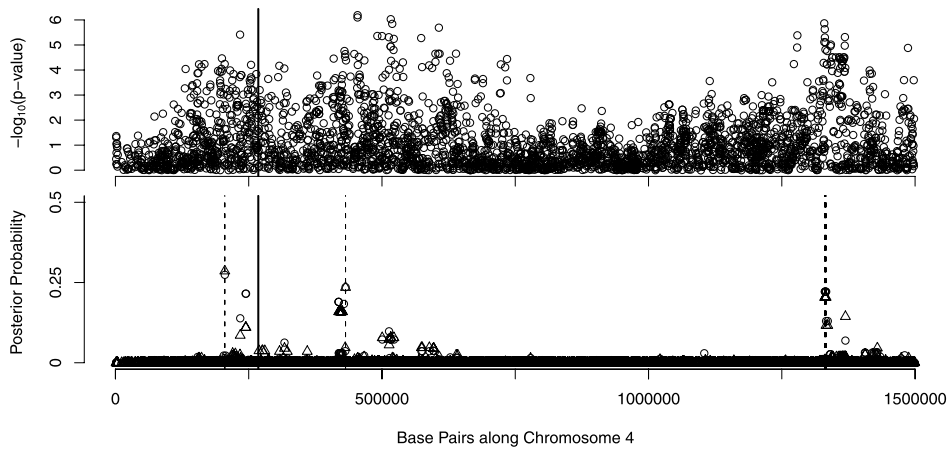}

\caption{Analysis of expression levels of \texttt{FLC}
for \textit{Arabidopsis thaliana}. The top plot shows results from
\textup{Single}, the bottom plot, which has a truncated $y$-axis,
shows results
from two runs of \textup{Sparse Partitioning}. Full details are
provided in the main text.}
\label{fig:flc}
\end{figure}

We examined how \textit{Sparse Partitioning} would deal with a problem
encountered in the 2010 project's most recent paper [\citet{atwell}].
The expression level of the \texttt{FLC} gene is known to be affected
by polymorphisms in the \texttt{FRIGIDA} region [\citet{johanson}; \citet{shindo}]. Atwell et al. performed a one-SNP-at-a-time
association study using \texttt{FLC} expression as the response. Its
analysis produced results similar to our analysis by \textit{Single},
shown in
the top plot of Figure~\ref{fig:flc}. While some SNPs within the
\texttt{FRIGIDA} region (which is marked by a vertical line) achieved
genome-wide significance, two stronger groups of associations were
detected approximately 200 kbp and 1~Mbp to the right. Prior knowledge
would suggest these downstream associations are spurious. When Atwell
et al. repeated the analysis, but this time including in the
regression model two alleles of the \texttt{FRIGIDA} gene known to
affect \texttt{FLC}, the downstream associations vanished, increasing
suspicion that they were false positives. For the rest of this section,
we assume this to be the case.

The project's latest data release provides typing for 214,553 SNPs
across five chromosomes. We picked the 3509 SNPs located within the
first 1.5 Mbp of Chromosome 4. As this subset was a biased selection
(e.g., it contained over two-thirds of those SNPs with marginal
$p$-values less than $10^{-4}$), it was necessary to reflect
this when choosing the prior probability of association for \textit{Sparse Partitioning}. In the event, we settled upon a prior
probability of~1 in 3500.

We used imputed data for this analysis, as the increased SNP density
allowed missing values to be inferred more reliably. Similar to the
analysis of Atwell et al., we decided to correct only for
relatedness, as discussions with members of the Nordborg group
convinced us that adjusting for population structure risked removing
too much true signal. The bottom plot of Figure~\ref{fig:flc} shows the
results of \textit{Sparse Partitioning}. The dashed vertical lines
indicate the three regions where our method found most evidence of
association. While two false positives remained, \textit{Sparse
Partitioning} gave greatest recognition to the \texttt{FRIGIDA} region,
identifying a possible association approximately 60 kbp upstream of the
gene.

In this example, we had knowledge of the true causal region, allowing
us to identify the false associations. The concern is that this example
is one of many, and that most times we will not know the correct
answer. In these cases, the best we can hope is that a method
acknowledges the true and false positives, but recognizes the
uncertainty. This is what \textit{Sparse Partitioning} has done here.
Furthermore, our method more precisely identified peaks than \textit{Single}
which should speed up the verification process.

\subsection{Mouse data}
\label{sub:mice}

Jon Krohn, from Professor Jonathan Flint's group at the University of
Oxford, kindly provided us with CD4 counts for 1274 ``heterogeneous
stock'' mice [\citet{solberg}], along with genotypic values for 770
SNPs covering the length of Chromosome 5. Krohn had previously analyzed
this data set using \textit{Bagphenotype}, software designed by Dr.
William Valdar (\url{http://www.unc.edu/\textasciitilde wvaldar/bagphenotype.html}).
The respon\-se values were continuous, while the predictors were
tertiary. Only a tiny proportion of genotypes (0.1\%) were missing, so
we saw no need to impute values and instead left them as unobserved. In
addition, we were provided with the gender of each mouse, which we
coded as a binary variable and included in the set of predictors.

\begin{figure*}

\includegraphics[scale=0.99]{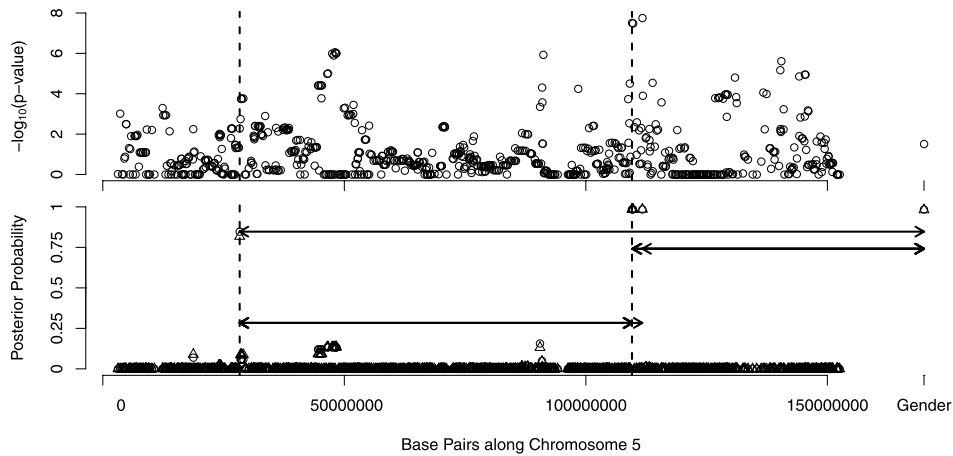}

\caption{Analysis of CD4 count in mice. The top plot
shows results from \textup{Single}, the bottom plot shows results from
two runs of \textup{Sparse Partitioning}. Full details are provided in the main
text.} \label{fig:mice}
\end{figure*}

As the chromosomal region was a subsection of a genome-wide study, we
decided a prior probability of association of 1 in 10,000 was
appropriate for each SNP. There is overwhelming prior knowledge that
CD4 counts are linked to gender [e.g., \citet{cd4count}], so we decided
upon a prior probability of 0.5. We run \textit{Sparse Partitioning}
allowing three copies of each predictor ($C=3$). As
Figure~\ref{fig:mice} demonstrates, the top hits from \textit
{Single}, SNPs
CEL-5\_106584673 and rs13478460, which due to linkage disequilibrium
are almost identical, persisted in \textit{Sparse Partitioning}. In
addition, our method declares associated SNP rs13478156. As indicated
by the horizontal arrows, \textit{Sparse Partitioning} found evidence
of interactions between gender and the top SNPs. To test the effect of
our prior choices, we repeated the analysis with prior probabilities
\{$10^{-4}$,~0.1\}, \{$10^{-3}$,~0.5\} and
\{$10^{-3}$,~0.1\}, and obtained very similar results on each
occasion (results not shown).

Maximum likelihood tests offer justification for why rs13478156 was
found by \textit{Sparse Partitioning}, but largely overlooked by
\textit{Single}. The supplementary material provides plots for two
sets of
tests. The first set compares, for each SNP, the pairwise interaction
with gender against a null model of no association. We find that the
top hits of \textit{Single} remain the most significant hits here. However,
these tests provide only limited information about the strength of the
interaction terms. Therefore, the second set compares the pairwise
interaction model against the additive model for that SNP and gender.
We see that the interaction between rs13478156 and gender is highly
significant. This supports \textit{Sparse Partitioning}'s claim that
this SNP acts in a gender specific way, which also agrees with findings
from Krohn's analysis.

This data set demonstrated the advantage of allowing multiple copies of
each predictor. When \textit{Sparse Partitioning} is run without this
option (results for $C=1$ are shown in the supplementary material), the
best fitting partition features three associated predictors in a single
nonnull group. The posterior estimates of pairwise interactions cannot be trusted because the method is unable to distinguish between,
say, a single three-way interaction and a pair of two-way interactions.
Allowing multiple copies of predictors requires only a small increase
in computation time, so we recommend this option is used.

\section{Discussion}
\label{sec:discussion}

It is fairly easy to design a regression method that is finely tuned
for a specific underlying relationship and then demonstrate its
superior power on data sets which obey this model. If one were
presented only with the results of the Model III simulations, it would
be easy to think that \textit{Sparse Partitioning} is such a method. We
have tried to show this is not case. We believe that \textit{Sparse
Partitioning} offers a robust alternative to existing methods. It fares
equally well under simple models, but comes into its own as the model
becomes more complex.

\subsection{Prospects for nonlinear regression}

Nonlinear regression methods~are competing over a fairly small share of
the market, bounded on the one side by the performance of methods such
as \textit{Single} and on the other by the strength of signal present
in the
data. Despite these limitations, there remains a demand for such
methods. There are many examples where standard linear methods fail to
explain a satisfactory percent of the variation, so it is quite
possible that nonadditive systems are at work. \textit{Sparse
Partitioning} should not be viewed as a search for interactions, but
rather as a regression method which bears interactions in mind. Even
for situations in which it cannot pinpoint an interaction with
certainty, its detection power should benefit for having considered
their existence.

\subsection{Generality}

\textit{Sparse Partitioning}'s strength derives from the generality of
its underlying relationship. Therefore, it is perhaps a surprise that
the method does not appear to suffer in situations where this
relationship is overly complicated. The results in
Section~\ref{sec:simstudy} suggest there is no inherent disadvantage to
using such a general underlying relationship. While \textit{Sparse
Partitioning} will almost certainly overfit the true model at some
points in the MCMC sampling, its posterior estimates are based on model
averages, rather than the single highest scoring model visited. For
this reason, it should not matter if a nonassociated predictor is
occasionally declared associated, as these errors are likely to be
spread thinly across the noncausal predictors. Additionally, as
\textit{Sparse Partitioning} seeks only to estimate marginal posterior
probabilities, using an underlying relationship too general should not
upset the Bayesian mechanics. The prior probability that a predictor is
associated remains constant (equal to $p_g$) regardless of the size of
the model space. Even if excessive generality does affect some aspects
of the posterior distribution, the marginal posterior probabilities
should remain correct.

\subsection{Diagnosis}

The only way to calculate the posterior distribution exactly is through
an exhaustive search of the space of partitions. Unfortunately, this is
feasible for only the smallest data sets, so instead \textit{Sparse
Partitioning} is forced to explore the model space in a stepwise
fashion. In this respect, \textit{Sparse Partitioning}'s search holds
an advantage over deterministic algorithms. When deciding which model
in the neighborhood to visit next, \textit{Sparse Partitioning} is not
forced to move to the highest scoring model. Instead it is able to try
a lower scoring move, in the hope that this is a gateway to a higher
scoring region.

The drawback of this stochasticity is the variability it introduces
into \textit{Sparse Partitioning}'s results. The analysis in
Section~\ref{sec:real} provides
some tips for gauging \textit{Sparse Partitioning}'s performance. It
is sensible to compare the
results with those of \textit{Single}, as we would expect very strong
associations to be found by both methods. Repeating the analysis with a
new random seed will highlight obvious lack of convergence, as should
examination of trace plots. Additionally, if time permits, repeating
the analysis with the response values permuted will provide
significance thresholds under a model of no true
associations.

\subsection{Limitations}

The processing time required for each iteration scales linearly with
$N$. We speculate that the number of iterations required for
convergence scales approximately with the 1.5th power of $N$ (based on
the stepwise nature of \textit{Sparse Partitioning}'s search) and
exponentially with the number
of true associations (based on the growth of the model
space).

As a rule of thumb, we consider \textit{Sparse Partitioning} suitable
for problems with no more than 20,000 predictors, or cases where
$N/n<100$. This is not to say that \textit{Sparse Partitioning} cannot
be applied on, say, a genome-wide scale, but it may be necessary to
filter the predictors first. We suggest picking, for example, the
highest 10\% of hits from \textit{Single}. Of course, this is not
ideal. It is
certainly possible that true associations are concealed within the
remaining 90\% of predictors. But considering standard practice
involves picking, perhaps, the top 100 hits of \textit{Single} for further
analysis, the ability to consider instead the top few thousand hits
should offer a significant advantage. As we experienced in
Section~\ref{sec:real}, it is important to realize when we have
selected a biased subset of predictors and reflect this in the prior
probability of association. The easiest solution is to pick the priors
as if analyzing the complete set of predictors.

We have identified situations in which \textit{Sparse Partitioning}
will struggle. Examples were found in Simulation Studies Five and Ten
(see Supplementary Material). The latter was an almost unavoidable
situation because the true relationship heavily contradicted our prior
beliefs. The former demonstrated the drawback of treating tertiary
predictors as categorical variables, when in fact their values have a
natural ordering. We suspect that this problem can be overcome by
application of a Bayesian version of Projection Pursuit [described in
\citet{esl}] that we are now developing.

Additionally, consider the case in which the response is influenced by
an interaction of two predictors, but the inclusion of neither
predictor on its own significantly improves the model fit. For one of
these predictors to have a realistic chance of being included in a
nonnull group, the improvement in fit must offset the penalty of
inclusion implied by $\mathbb{P}(\mathbb{G})$. Because of the single-step
nature of \textit{Sparse Partitioning}'s search, it is unlikely that
either predictor will
appear in the current model, which is required for the method to
consider their interaction. For our method to be successful in this
case, it would have to permit two-step moves or resort to an exhaustive
search.\looseness=1

\textit{Sparse Partitioning} can be used when the predictors are
continuous, provided a suitable transformation exists. For example, we
have applied our method to copy number values, by first reducing each
continuous measurement to one of three classes (neutral, increased or
decreased). In the same way, we hope our method can be applied to a
whole range of problems.\looseness=1

\begin{appendix}\label{app}

\section*{Appendix: Details of Bayesian Framework}
\label{app:method}

The regression model is written as $l(\mathbb{E}(\mathbf{Y})) =
f(\mathbf{X})$, where $l$ is a specified link function and
$f(\mathbf{X})$ is the ``underlying relationship.'' Without loss of
generality, the underlying relationship can be expressed as the sum of
functions of groups of associated predictors:
\[
f(\mathbf{X}) = f_1({X}_{\mathbf{G}_1}) + f_2({X}_{\mathbf{G}_2}) +
\cdots+f_K({X}_{\mathbf{G}_K}).
\]
The disjoint sets $\mathbf{G}_1$, $\mathbf{G}_2, \ldots, \mathbf{G}_K$ index groups
of associated predictors. Let $\mathbf{G}_0$, the ``null group,'' index the
predictors not associated. Therefore, $\mathbb{G}= \{\mathbf{G}_0,\mathbf{G}_1,\mathbf{G}_2, \ldots, \mathbf{G}_K\}$ partitions $\{1,2,\ldots,N\}$.
Equivalently, the partition can be described by the vector $\mathbf{I} =
(I_1, I_2,\ldots,I_N)$, where $I_g$ indicates to which group the $g$th
predictor belongs. Only unique partitions are considered, so the
ordering of elements within groups is irrelevant, as is the ordering of
nonnull groups.

A single model will be $\{\mathbb{G}, \mathbf{f}\}$, a partition and a
corresponding set of functions $\{f_1,f_2, \ldots, f_K\}$. The model
space will be all such permissible pairs. If we wish to allow
predictors to feature in more than one group of associations, the
predictor set is expanded to contain $C$ copies of each predictor. An
alternative approach is to keep one copy of each predictor, but relax
the condition on disjoint groups. However, we felt this approach
created a~greater amount of duplication within the space of underlying
relationships, making it more challenging to define a prior. The
description of the method supposes $C=1$, with the alternative case
discussed when necessary.

We are interested in the posterior distribution of $\mathbb{G}$ and
$\mathbf{f}$, given the observed values for $\mathbf{X}$ and $\mathbf{Y}$.
To be fully Bayesian, we must also consider the distribution of the
predictors, which can be written as $\mathbb{P}(\mathbf{X}|\varepsilon)$,
for some parameter vector $\varepsilon$:
\[
\mathbb{P}(\mathbb{G}, \mathbf{f}, \varepsilon| \mathbf{X}, \mathbf{Y})
\propto\mathbb{P}(\mathbb{G}, \mathbf{f}, | \mathbf{X}, \mathbf{Y})
\times\mathbb{P} (\varepsilon| \mathbb{G}, \mathbf{f}, \mathbf{X}, \mathbf{Y}).
\]
If we assume $\varepsilon$ is unaffected by $\mathbb{G}$ and $\mathbf{f}$
[\citet{gelman}], its posterior can be ignored in the calculation of
$\mathbb{P}(\mathbb{G},\mathbf{f} | \mathbf{X}, \mathbf{Y})$. Similarly, as
we only wish to estimate properties of the posterior distribution of
partitions, we treat the functions as nuisance parameters:
\begin{eqnarray*}
\mathbb{P}(\mathbb{G}|\mathbf{X}, \mathbf{Y})
&\propto&
\mathbb{P}(\mathbb {G}|\mathbf{X})\times\mathbb{P}(\mathbf{Y}| \mathbf{X}, \mathbb{G})
\\[-2pt]
&=&
\mathbb{P}(\mathbb{G}|\mathbf{X})\times\int_{\mathbf{f}}\mathbb{P}(\mathbf{Y}|\mathbf{f},\mathbf{X},
\mathbb{G})\mathbb{P}(\mathbf{f}|\mathbf{X},\mathbb{G})\,d\mathbf{f},
\end{eqnarray*}
with $\mathbb{P}(\mathbb{G}|\mathbf{X})$ reducing to
$\mathbb{P}(\mathbb {G})$, as we suppose the prior distribution of
$\mathbb{G}$ does not depend on the observed values of the
predictors.

\subsection{Partition prior, $\mathbb{P}(\mathbb{G})$}

The prior for the partition is constructed so the probability that
predictor $g$ is associated equals $p_g$. For partition $\mathbf{I}$, we can
define the equivalence class $[\mathbf{I}]$ containing all partitions that
declare\vadjust{\eject} the same predictors associated. To ensure the marginal
probability that predictor $g$ is associated equals $p_g$, we desire
\[
\mathbb{P}([\mathbf{I}]) = \sum_{\mathbf{I}' \in[\mathbf{I}]} \mathbb{P}(\mathbf{I}') =
\prod_{j:I_j=0} (1-p_j) \prod_{j:I_j\neq0} p_j = \prod_{j \in\mathbf{G}_0} (1-p_j) \prod_{j\notin\mathbf{G}_0} p_j,
\]
because then
\begin{eqnarray*}
\mathbb{P}(I_g\neq0)
&=&
\sum_{\mathbf{I}: I_g \neq0}\mathbb{P}(\mathbf{I}) =
\sum
_{[\mathbf{I}]:I_g\neq0} \biggl( \prod_{j:I_j=0} (1-p_j) \prod_{j:I_j\neq0} p_j \biggr)
\\
&=&
p_g \prod_{j\neq g} [(1-p_j)+p_j],
\end{eqnarray*}
equaling $p_g$, as required.

Assigning equal weighting to members of $[\mathbf{I}]$, we can calculate
$\mathbb{P}(\mathbf{I})$ explicitly by counting the size of each equivalence
class. If $\mathbf{I}$ declares $s=N-|\mathbf{G}_0|$ predictors associated, then
the size of $[\mathbf{I}]$ will be the number of ways $s$ elements can be
partitioned. Unrestricted, this would equal the $s$th Bell number.
Instead, \textit{Sparse Partitioning} limits each partition to no more
than $K$ nonnull groups, each containing at most $S$ elements. These
``truncated'' Bell numbers, $B(s,K,S)$, can be calculated in a
recursive fashion. Let $a_j$ denote the number of groups of size $j$
for $j=1,2,\ldots,S$. Then
\begin{eqnarray*}
&&
B(s,K,S\vert a_1,a_2,\ldots,a_{S-1},a_S)
\\
&&\qquad =
B(s,K,S\vert a_1-1,a_2,\ldots,a_{S-1},a_S)
\\
&&\qquad\quad{} +
B(s,K,S\vert a_1+1,a_2-1,\ldots,a_{S-1},a_S)
(a_1+1)
 \\
&&\qquad\quad{}+ \cdots+
B(s,K,S\vert a_1,a_2,\ldots,a_{S-1}+1,a_S-1)
(a_{S-1}+1),
\end{eqnarray*}
with boundary condition
\[
B(0,K,S|a_1,a_2,\ldots,a_{S-1},a_S)=
\cases{
1,&\quad if $a_j= 0$ $\forall j$,\cr
0,&\quad otherwise.
}
\]
Equally weighting each member of $[\mathbf{I}]$ places a high probability
($1-|[\mathbf{I}]|^{-1}$) on the existence of interactions, even though few
interactions have so far been found and verified. It would be
straightforward to alter the partition weightings. For example, we
could choose to favor partitions containing fewer interactions.
However, the lack of known interactions must largely be due to how hard
they are to identify, coupled with how rarely they are searched for.
Therefore, we are satisfied that a uniform weighting is a reasonable
choice.

\textit{Sparse Partitioning} requires that $K$ and $S$ are set in
advance, to allow sufficient memory to be allocated and pre-calculation
of $B(s,K,S)$. Theoretically, $K$ and $S$ should be no smaller than
$N$, to ensure the two most extreme underlying relationships are
possible (either $N$ groups of size one or one group of size $N$). In
practice, these values would require vast amounts of unnecessary
computation. Therefore, we suggest $K$ and $S$ are set to the smallest
values possible, without impacting the direction of the MCMC chain.

The calculation of $\mathbb{P}(I_g\neq0)$ assumes $K\times S \geq N$, as
the last summation supposes all $2^N$ equivalence classes are
achievable. When this condition does not hold, the error involved can
be calculated for the case that all prior probabilities are equal:
\begin{eqnarray*}
\mathbb{P}(I_g\neq0)
&=&
p_g \sum_{s=0}^{KS-1} \pmatrix{N-1\cr s} p_g^s
(1-p_g)^{N-1-s}\biggl/\sum_{s=0}^{KS} \pmatrix{N\cr s} p^s_g (1-p_g)^{N-s}
\\
&=&
p_g \mathbb{P}\bigl(s\leq KS-1\mid s\sim\mathbb{B}(p_g,N-1)\bigr)/\mathbb
{P}\bigl(s\leq
KS\mid s\sim\mathbb{B}(p_g,N)\bigr).
\end{eqnarray*}

Using a normal approximation for each binomially distributed variable,
we obtain
\[
\mathbb{P}(I_g\neq0) = p_g \Phi\biggl(
\frac{KS-1/2-(N-1)p_g}{\sqrt{p_g(1-p_g)(N-1)}}\biggr) \biggl/\Phi\biggl(
\frac{KS+1/2-Np_g}{\sqrt{p_g(1-p_g)N}}\biggr),
\]
where $\Phi$ is the cumulative probability function for a standard
normal. For small $p_g$, the value of $\mathbb{P}(I_g\neq0)$ is affected
most by the prior mean, $Np_g$. We suggest setting $K=4$ and $S=4$.
Entering these values into the equation above, we find that the actual
prior probability of association used by \textit{Sparse Partitioning}
lies within 1\% of the desired value, $p_g$, even when the prior mean
is as high as 9.

When multiple copies of predictors are allowed ($C > 1$), the prior
probability of association for each copy of predictor $g$ is set to
$1-\sqrt[C]{(1-p_g)}$. This ensures the probability that one or more
copies of predictor $g$ are associated remains equal to $p_g$. Allowing
multiple copies of predictors creates an element of duplication within
the space of partitions. For example, a partition in which two copies
of a predictor feature in the same nonnull group effects the same
underlying relationship as the partition obtained when one of these
copies is removed. As a result, the prior weighting for this underlying
relationship is increased. However, for small values of $p_g$ this
effect will be negligible. As with $K$ and $S$, it is necessary to
specify $C$ in advance. Its value has minimal effect on computation
time, so we recommend a conservative setting, such as $C=3$.

\subsection{Function prior, $\mathbb{P}(\mathbf{f}|\mathbb{G})$}

To ensure identifiability of the functions, one value of $X_{\mathbf{G}_k}$
is considered the base value and its mapping is absorbed into the
overall intercept (denoted by $\alpha_0$). Therefore, $f_k$ has degree
of freedom one less than $d_k$, the number of unique values (nodes) of
$X_{\mathbf{G}_k}$. Let $V_{k,1}, V_{k,2}, \ldots,V_{k,d_k-1}$ be dummy
binary variables that distinguish the remaining $d_k-1$ nodes; these
map to $\alpha_{k,1}, \alpha_{k,2}, \ldots, \alpha_{k,d_k-1}$,
respectively. The underlying relationship can be written in standard
regression form:
\begin{eqnarray*}
f(\mathbf{X})
&=&
\alpha_0 + ( \alpha_{1,1} V_{1,1} + \cdots+\alpha
_{1,{d_1-1}} V_{1,{d_1-1}} )
\\
&&{}+ \cdots+
(\alpha_{K,1} V_{K,1} + \cdots+ \alpha_{K,{d_K-1}} V_{K,{d_K-1}}).
\end{eqnarray*}
All the relevant information of the functions is contained in the
vector $\bolds{\alpha}= \{\alpha_0,\alpha_{1,1}, \ldots,
\alpha_{1,d_1-1},\ldots,\alpha_{K,1}, \ldots, \alpha_{K,d_K-1}\}$, of
size $D=1+\sum(d_k-1)$. \textit{Sparse Partitioning} assigns
independent normal priors with mean 0 to each element of $\bolds{\alpha}$.
These can be viewed as a penalty on smoothness, but one which accepts
that with categorical predictors there is no ordering to the nodes.
This agrees with a belief in parsimony, which prefers simple functions
to complicated ones.

In the continuous response case the variance of these normal priors is
$\sigma^2 /r$; in the binary response case the variance is $1/r$. In
both cases, the choice of $r$ controls the extent by which smoothness
is applied. Typically we set $r$ to 1.

\subsection{Likelihood, $\mathbb{P}(\mathbf{Y}|\mathbf{f}, \mathbf{X},\mathbb{G})$}

When the response is continuous, the link function is the identity and
the residuals are assumed to be independent draws from a normal
distribution with mean zero and variance $\sigma^2$:
\begin{eqnarray*}
\mathbb{P}(\mathbf{Y}|\mathbf{f}, \mathbf{X}, \mathbb{G})
&=&
\int_{\sigma^2}\mathbb{P} (\mathbf{Y}|\sigma^2, \mathbf{f}, \mathbf{X}, \mathbb{G})\mathbb{P}(\sigma^2)\,d\sigma^2
\\
&=&
\int_{\sigma^2}(2\pi\sigma^2)^{-n/2} \exp\biggl\{-\frac{1}{2\sigma^2}\bigl(\mathbf{Y}-f(\mathbf{X})\bigr)^T\bigl(\mathbf{Y}-f(\mathbf{X})\bigr)\biggr\} \sigma^{-2}\,d\sigma^2.
\end{eqnarray*}
This integral incorporates a prior for $\sigma^2$ of the form
$\sigma^{-2}$, which reflects a~preference for smaller variances. It
does not matter that this prior is improper as it is common to all
models.

When the response is binary, a logit link function is used, $l(a) =
\log(\frac{a}{1-a})$:
\[
\mathbb{P}(\mathbf{Y}|\mathbf{f}, \mathbf{X}, \mathbb{G}) =
\prod_{i} [l^{-1}f(X_{i.})]^{ Y_i}
[1-l^{-1}f(X_{i.})]^{(1-Y_i)}.
\]

\subsection{Marginal likelihood, $\mathbb{P}(\mathbf{Y}|\mathbf{X}, \mathbb{G})$}

\begin{eqnarray*}
\mathbb{P}(\mathbf{Y}|\mathbf{X}, \mathbb{G})
&=&
\int_{\mathbf{f}} \mathbb{P}(\mathbf{Y}|\mathbf{f}, \mathbf{X}, \mathbb{G})\mathbb{P}(\mathbf{f}| \mathbf{X}, \mathbb{G})\,d\mathbf{f}
\\
&=&
\int_{\bolds{\alpha}} \mathbb{P}(\mathbf{Y}| \bolds{\alpha}, \mathbf{X},\mathbb{G}) \mathbb{P}(\bolds{\alpha}|\mathbf{X}, \mathbb{G})\,d\bolds{\alpha}.
\end{eqnarray*}
With a continuous response, $\mathbb{P}(\mathbf{Y} | \mathbf{X},
\mathbb{G})$ can be calculated explicitly. When the response is binary,
\textit{Sparse Partitioning} uses a Laplace approximation. Let
$W(\bolds{\alpha}) = \mathbb{P}(\mathbf{Y}|\bolds{\alpha}, \mathbf{X},
\mathbb{G}) \mathbb{P}(\bolds{\alpha}|\mathbf{X}, \mathbb{G})$ and
$w(\bolds{\alpha}) =\log(W(\bolds{\alpha}))$:
\[
w(\bolds{\alpha}) \approx w(\bolds{\alpha}') + (\bolds{\alpha}- \bolds{\alpha}')^T
\frac{dw(\bolds{\alpha}')}{d\bolds{\alpha}} + \frac{1}{2}(\bolds{\alpha}- \bolds{\alpha}')^T \frac{d^2w(\bolds{\alpha}')}{d\bolds{\alpha}^2} (\bolds{\alpha}-
\bolds{\alpha}').
\]
If $\hat{\bolds{\alpha}}$ is the maximum likelihood estimate of $w(\bolds{\alpha})$, then
\[
W(\bolds{\alpha}) \approx W(\hat{\bolds{\alpha}}) \exp\biggl\{ -\frac{1}{2}(\bolds{\alpha}-
\hat{\bolds{\alpha}})^T \biggl( -\frac{d^2w(\hat{\bolds{\alpha}})}{d\bolds{\alpha}
^2} \biggr)
(\bolds{\alpha}- \hat{\bolds{\alpha}}) \biggr\}.
\]
Therefore,
\[
\mathbb{P}(\mathbf{Y}|\mathbf{X}, \mathbb{G})
\approx\mathbb{P}(\mathbf{Y}| \hat{\bolds\alpha}, \mathbf{X}, \mathbb{G})
\mathbb{P}(\hat{\bolds\alpha}|\mathbf{X}, \mathbb{G}) (2\pi)^{D/2}
\biggl| -\frac{d^2w(\hat{\bolds\alpha})}{d\bolds{\alpha} ^2} \biggr| ^{-1/2}.
\]
Alternatively, \textit{Sparse Partitioning} allows the user to select a
probit link function, in which case a latent variable representation of
the likelihood can be used [\citet{albert}]. Essentially, each binary
response is replaced by a continuous ``pseudo response.'' The
regression model is then treated as if it were linear, except the new
response values are resampled once per iteration.

When there are two or more functions present, the marginal likelihood
will be affected (very slightly) by which node is considered the base
value for each function. For consistency, the node removed is chosen
according to a defined rule (and is the zero vector of $X_{\mathbf{G}_k}$ if
available). In addition, before analysis begins, continuous response
values are transformed to have mean 0 and variance 1 to reduce
variability caused by the choice of base value.
\end{appendix}

\section*{Software} \textit{Sparse Partitioning} has been
implemented and is available at
\url{http://www.compbio.group.cam.ac.uk/software.html}.

\section*{Acknowledgments}
The authors thank Antigone Dimas, Keyan Zhao, Bjarni Vilhj\'{a}lmsson
and Jon Krohn for providing data from their respective laboratories. We
thank Alexandra Jauhiainen, Terry Speed, Michael Stein and an anonymous
reviewer for helpful suggestions on earlier versions of this article,
and we acknowledge the support of The University of Cambridge, Cancer
Research UK and Hutchison Whampoa Limited.

\begin{supplement}[id=supplement]
\sname{Supplement}
\stitle{Extra material}
\slink[doi]{10.1214/10-AOAS411SUPP}
\slink[url]{http://lib.stat.cmu.edu/aoas/411/supplement.pdf}
\sdatatype{.pdf}
\sdescription{Provides additional details of \textit{Sparse
Partitioning}'s methodology, full
explanation of the simulation studies and extended results from
applying the method to real data sets.}

\end{supplement}


\printaddresses


\begin{thebibliography}{99}

\bibitem[\protect\citeauthoryear{Albert and Chib}{1993}]{albert}
\textsc{Albert, J.} and \textsc{Chib, S.} (1993). Bayesian analysis of
binary and polychotomous response data. \textit{J.~Amer. Statist.
Assoc.} \textbf{88} 669--679.
\MR{1224394}

\bibitem[\protect\citeauthoryear{Atwell et~al.}{2010}]{atwell}
\textsc{Atwell, S., Huang, Y., Vilhj{\'{a}}lmsson, B., Willems, G.,
Horton, M.} and \textsc{Li, Y.} (2010). Genome-wide association study
of 107 phenotypes in \textit{Arabidopsis thaliana} inbred lines.
\textit{Nature} \textbf{465} 627--631.

\bibitem[\protect\citeauthoryear{Balding}{2006}]{balding}
\textsc{Balding, D.} (2006). A tutorial on statistical methods for
population association studies. \textit{Nat. Rev. Genet.} \textbf{7}
781--791.

\bibitem[\protect\citeauthoryear{Breiman}{2004}]{rf}
\textsc{Breiman, L.} (2004). Random Forests. \textit{Machine Learning}
\textbf{45} 5--32.

\bibitem[\protect\citeauthoryear{Breiman et~al.}{1984}]{carttrees}
\textsc{Breiman, L., Friedman, J., Olshen, R.} and \textsc{Stone, C.}
(1984). \textit{Classification and Regression Trees}. Wadsworth,
Belmont, CA.
\MR{0726392}

\bibitem[\protect\citeauthoryear{Cordell}{2009}]{cordell}
\textsc{Cordell, H.} (2009). Detecting gene-gene interactions that
underlie human diseases. \textit{Nat. Rev. Genet.} \textbf{10}
392--404.

\bibitem[\protect\citeauthoryear{Dimas}{2009}]{dimas}
\textsc{Dimas, A.} (2009). The role of regulatory variation in
sculpting gene expression across human populations and cell types. Ph.D.
thesis, Darwin College, Univ. Cambridge.

\bibitem[\protect\citeauthoryear{Gelman et~al.}{2004}]{gelman}
\textsc{Gelman, A., Carlin, J., Stern, H.} and \textsc{Rubin, D.}
(2004). \textit{Bayesian Data Analysis}. Chapman and Hall/CRC, Boca
Raton, FL.
\MR{2027492}

\bibitem[\protect\citeauthoryear{Hans, Dobra and West}{2007}]{sss}
\textsc{Hans, C., Dobra, A.} and \textsc{West, M.} (2007). Shotgun
stochastic search for ``large $p$'' regression. \textit{J.~Amer.
Statist. Assoc.} \textbf{102} 507--516.
\MR{2370849}

\bibitem[\protect\citeauthoryear{Hastie, Tibshirani and Friedman}{2001}]{esl}
\textsc{Hastie, T., Tibshirani, R.} and \textsc{Friedman, J.} (2001).
\textit{The Elements of Statistical Learning}. Springer, New York.
\MR{1851606}

\bibitem[\protect\citeauthoryear{Hoggart et~al.}{2008}]{hoggart}
\textsc{Hoggart, C., Whittaker, J., De Iorio, M.} and \textsc{Balding,
D.} (2008). Simultaneous analysis of all {SNPs} in genome-wide and
re-sequencing association studies. \textit{PLoS Genet.} \textbf{4}
e10000130.

\bibitem[\protect\citeauthoryear{Johanson et~al.}{2000}]{johanson}
\textsc{Johanson, U., West, J., Lister, C., Michaels, S., Amasino, R.}
and \textsc{Dean, C.} (2000). Molecular analysis of FRIGIDA, a major
determinant of natural variation in \textit{Arabidopsis} flowering
time. \textit{Science} \textbf{290} 344--347.

\bibitem[\protect\citeauthoryear{Maini et~al.}{1996}]{cd4count}
\textsc{Maini, M., Gilson, R., Chavda, N., Gill, S., Fakoya, A., Ross,
E., Phillips, A.} and \textsc{Weller, I.} (1996). Reference ranges and
sources of variability of {CD4} counts in HIV-seronegative women and
men. \textit{Genitourin. Med.} \textbf{72} 27--31.

\bibitem[\protect\citeauthoryear{Marchini, Donnelly and
Cardon}{2005}]{marchini}
\textsc{Marchini, J., Donnelly, P.} and \textsc{Cardon, L.} (2005).
Genome-wide strategies for detecting multiple loci that influence
complex diseases. \textit{Nat. Genet.} \textbf{37} 413--417.

\bibitem[\protect\citeauthoryear{{McC}arthy et~al.}{2008}]{mccarthy}
\textsc{McCarthy, M., Abecasis, G., Cardon, L., Goldstein, D., Little,
J., Ioannidis,~J.} and \textsc{Hirschhorn, J.} (2008). Genome-wide
association studies for complex traits: Consensus, uncertainty and
challenges. \textit{Nat. Rev. Genet.} \textbf{10} 356--369.

\bibitem[\protect\citeauthoryear{Ruczinski, Kooperberg and
LeBlanc}{2003}]{logic}
\textsc{Ruczinski, I., Kooperberg, C.} and \textsc{LeBlanc, M.} (2003).
Logic regression. \textit{J. Comput. Graph. Stat.} \textbf{12}
475--511.
\MR{2002632}

\bibitem[\protect\citeauthoryear{Shindo et~al.}{2005}]{shindo}
\textsc{Shindo, C., Aranzana, M., Lister, C., Baxter, C., Nicholls, C.,
Nordborg, M.} and \textsc{Dean, C.} (2005). Role of FRIGIDA and
FLOWERING LOCUS C in determining variation in flowering time of
\textit{Arabidopsis thaliana. Plant Physiol.} \textbf{138} 1163--1173.

\bibitem[\protect\citeauthoryear{Solberg et~al.}{2006}]{solberg}
\textsc{Solberg, L., Valdar, W., Gauguier, D., Nunez, G., Taylor, A.,
Burnett, S., Arboledas-Hita, C., Hernandez-Pliego, P., Davidson, S.,
Burns, P., Bhattacharya, S., Hough, T., Higgs, D., Klenerman, P.,
Cookson, W., Zhang,~Y., Deacon, R., Rawlins, J., Mott, R.} and
\textsc{Flint, J.} (2006). A protocol for high-throughput phenotyping,
suitable for quantitative trait analysis in mice. \textit{Mamm. Genome}
\textbf{17} 129--146.

\bibitem[\protect\citeauthoryear{Speed and Tavar{\'e}}{2010}]{speedsupp}
\textsc{Speed, D.} and \textsc{Tavar{\'e}, S.} (2010). Supplement to
``Sparse {P}artitioning: Nonlinear regression with binary or tertiary
predictors with application to association studies.'' DOI:
\href{http://dx.doi.org/10.1214/10-AOAS411SUPP}{10.1214/10-AOAS411SUPP}.

\bibitem[\protect\citeauthoryear{Stephens and Balding}{2009}]{stephbald}
\textsc{Stephens, M.} and \textsc{Balding, D.} (2009). Bayesian
statistical methods for genetic association studies. \textit{Nat. Rev.
Genet.} \textbf{10} 681--690.

\bibitem[\protect\citeauthoryear{Stranger et~al.}{2007}]{stranger}
\textsc{Stranger, B., Forrest, M., Dunning, M., Ingle, C., Beazley, C.}
and \textsc{Thorne, N.} (2007). Relative impact of nucleotide and copy
number variation on gene expression phenotypes. \textit{Science}
\textbf{315} 848--853.

\bibitem[\protect\citeauthoryear{{The Wellcome Trust Case Control
Consortium}}{2007}]{wellcome}
\textsc{The Wellcome Trust Case Control Consortium} (2007).
Genome-wide association study of 14,000 cases of seven common diseases
and 3,000 shared controls. \textit{Nature} \textbf{447} 661--678.

\bibitem[\protect\citeauthoryear{Wang et~al.}{2005}]{wang}
\textsc{Wang, H., Zhang, Y., Li, X., Masinde, G., Mohan, S., Baylink,
D.} and \textsc{Xu, S.} (2005). Bayesian shrinkage estimation of
quantitative trait loci parameters. \textit{Genetics} \textbf{170}
465--480.

\bibitem[\protect\citeauthoryear{Zhang et~al.}{2005}]{zhang}
\textsc{Zhang, M., Montooth, K., Wells, M., Clark, A.} and
\textsc{Zhang, D.} (2005). Mapping multiple quantitative trait loci by
{B}ayesian classification. \textit{Genetics} \textbf{169} 2305--2318.

\bibitem[\protect\citeauthoryear{Zhao et~al.}{2007}]{zhao}
\textsc{Zhao, K., Aranzana, M., Kim, S., Lister, C., Shindo, C., Tang,
C., Toomajian,~C., Zheng, H., Dean, C., Marjoram, P.} and
\textsc{Nordborg, M.} (2007). An \textit{Arabidopsis} example of
association mapping in structured samples. \textit{PLoS Genet.}
\textbf{3} e4.

\end{thebibliography}
\end{document}